# Anomalous Freezing of Low Dimensional Water Confined in Graphene Nanowrinkles

*Tim Verhagen, [1#] Jiri Klimes, [2#] Barbara Pacakova,[3,4] Martin Kalbac, [3]\* and Jana Vejpravova[1]\**

[1]Department of Condensed Matter Physics, Charles University, Ke Karlovu 5, 121 16 – Prague 2, Czech Republic

[2]Department of Chemical Physics and Optics, Charles University, Ke Karlovu 3, 121 16 – Prague 2, Czech Republic

[3]JH Institute of Physical Chemistry, Dolejskova 3, 182 23 Prague 8, Czech Republic

[4]Norwegian University of Science and Technology, Department of Physics, Høgskoleringen 5, NO-7491, Trondheim, Norway

[#]These authors contributed equally to this work.

CORRESPONDING AUTHORS

\*martin.kalbac@jh-inst.cas.cz; jana@mag.mff.cuni.cz

ABSTRACT

Various properties of water are affected by confinement as the space-filling of the water molecules is very different from bulk water. In our study, we challenged the creation of a stable system in which water molecules are permanently locked in nanodimensional graphene traps. For that purpose, we developed a technique – nitrocellulose-assisted transfer of graphene grown by chemical vapor deposition, which enables capturing of the water molecules below an atomically thin graphene membrane structured into a net of regular wrinkles with a lateral dimension of about 4 nm. After successfully confining water molecules below a graphene monolayer, we employed cryogenic Raman spectroscopy to monitor the phase changes of the confined water as a function of the temperature. In our experiment system, the graphene monolayer structured into a net of fine wrinkles plays a dual role: i) it enables water confinement and ii) serves as an extremely sensitive probe for phase transitions involving water *via* graphene-based spectroscopic monitoring of the underlying water structure. Experimental findings were supported with classical and path integral molecular dynamics simulations carried out on our experiment system. Results of simulations show that surface premelting of the ice confined within the wrinkles starts at ~ 200 K and the melting process is complete at ~ 240 K, which is far below the melting temperature of bulk water ice. The processes correspond to changes in the doping and strain in the graphene tracked by Raman spectroscopy. We conclude that water can be confined between graphene structured into nanowrinkles and silica substrate and its phase transitions can be tracked *via* Raman spectral feature of the encapsulating graphene. Our study also demonstrated that peculiar behavior of liquids under spatial confinement can be inspected *via* the optical response of atomically thin graphene sensors.

TOC GRAPHICS

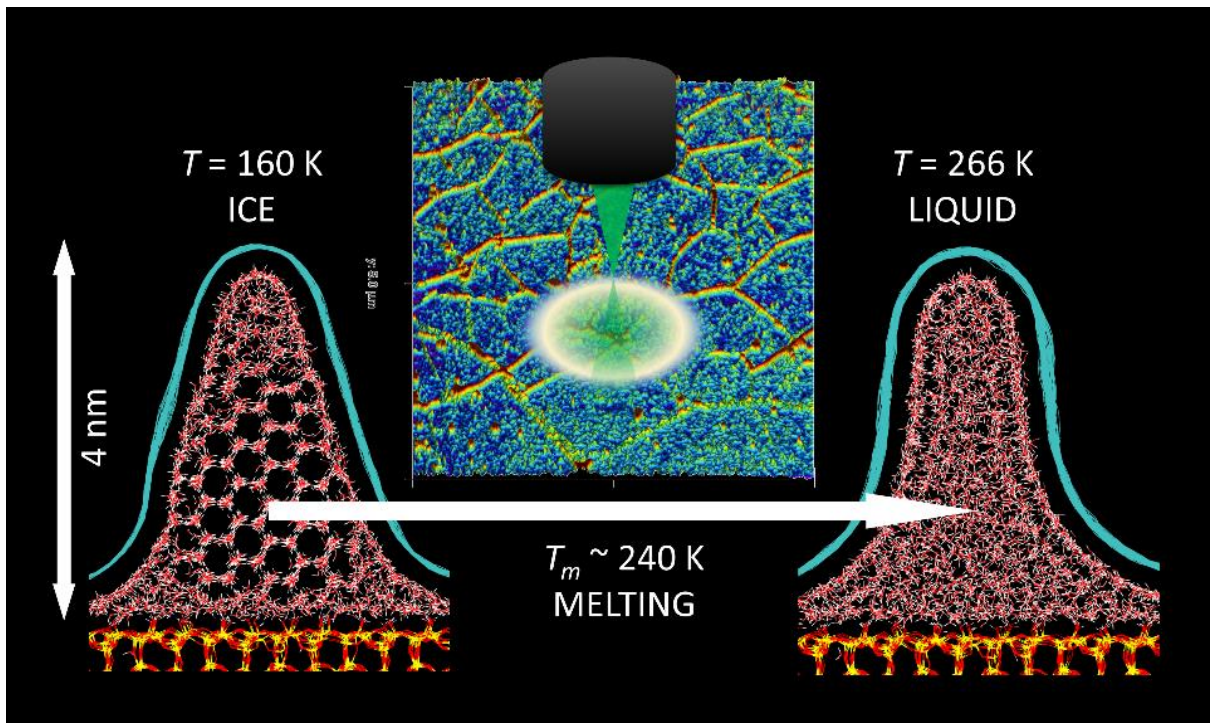

KEYWORDS

graphene, water confinement, ice melting, molecular dynamics simulations, cryogenic Raman spectroscopy, wrinkles

Graphene is a material of great promise due to its single-atom thickness and exceptional electronic structure. For example, its high sensitivity to adsorbed molecules could make it an ideal gas sensor.[1,2] However, its prompt and substantial response to any physiochemical stimuli can be also problematic for graphene-based devices as the charge carrier concentration is affected by water and thus changes due to ambient humidity variations.[3] In some cases water can even enter the interface between support and graphene layer, affecting substantially the structural and electronic properties of graphene.[4] As water is ubiquitous, such effects can occur unexpectedly and the presence of water has to be considered while analyzing results.

The effects of water entering the interface between substrate and graphene can be also turned into advantage by using them to understand some of the peculiar behavior of water. It is known that various properties of water are affected by confinement. For example, both increase and decrease of the melting temperature was observed for water confined in carbon nanotubes of different diameter.[5–7] Simulations showed a surface premelting and drop in melting temperature for nanometer-sized water ice clusters[8,9] and similar effects were observed for water confined in porous materials.[10] This even lead to development of confined-water phase diagram.[11] For water restricted between two graphene sheets, different two-dimensional ice phases have been predicted and/or observed.[12] It is not clear if such phases will still occur when less homogeneous substrate is used, such as silica. Moreover, the graphene layer is not typically flat but corrugated, possibly even with some regions delaminated from the substrate, making folds or wrinkles.[13] The thickness of water be can be thus different across the surface, leading to different importance of the various confining effects.

In a broader context, the liquids confined in nanometer scale exist in a plethora of systems ranging from mesoporous materials to interior of tissues and cells. As extensively discussed by Thompson,[14] the confinement plays a crucial role in understanding the connection between the molecular-level characteristics and the macroscopic behavior of the materials and natural systems. This area has received significant attention in recent years. In this perspective, the advances in understanding the dynamics of confined liquids remains the nontrivial challenge.

In this work we report anomalous behavior of water confined between silica substrate and monolayer graphene structured into a net of nanoscale wrinkles. We take advantage of the high sensitivity of graphene to water and use cryogenic Raman spectroscopy to track the changes in strain and doping of graphene. Unexpected behavior in doping and strain clearly points to the large effect of confined water on these properties. To understand the observations we performed extensive molecular dynamics (MD) and path integral molecular dynamics (PIMD) simulations. The simulations show that premelting of the confined water starts at around 200 K and water in graphene wrinkles melts at around 240 K, much below the melting temperature of bulk water. The observed changes in water dynamics and structure of graphene clearly correspond to the experimental data. The complex behavior of water confined between graphene and various substrates as well as the interplay between the confining material and the confined water can be effectively studied by this approach. Thus the use of the confining material as an *in situ* spectroscopic tool is a straightforward way to monitor volume changes and phase transitions induced by the temperature or pressure variations. Besides, we can address various physio-chemical processes and chemical reactions accommodated in the nanoscale graphene wrinkles.

RESULTS/DISCUSSION

In our study, a few layers of water were captured between the graphene monolayer grown by chemical vapor deposition (CVD) and $SiO_2$ (300 nm)/Si substrate. In the process, nitrocellulose polymer was deposited on top of the CVD-grown graphene monolayer as a transfer medium. More technical details on the wet nitrocellulose-assisted transfer process of the CVD-grown graphene can be found in previously published works[15,16] and here in the section "Methods/Experimental". The water molecules remained trapped between the graphene and the $SiO_2$/Si substrate as the transfer takes place in presence of water as a solvent. The nitrocellulose-assisted transfer yields a specific topography of the transferred graphene layer, consisting of a network of nanoscale wrinkles with a typical height of approximately 4 nm: the topography obtained by atomic force microscopy (AFM) and a schematic side view of a wrinkle are shown in Fig. 1(a) and Fig. 1(b), respectively. The parts of the graphene layer adhered to the substrate are denoted as 'flat', whereas the curved regions are called 'wrinkled'. We would like to stress that the final shape of the graphene wrinkles is determined by the water molecules trapped between the graphene layer and the $SiO_2$/Si substrate. The monolayer graphene also serves as a barrier for water molecules captured between the graphene layer and the substrate. We thus demonstrated that using several hundreds of nanometers large graphene membrane structured into a regular net of nanoscale wrinkles enables highly efficient confining water molecules.

The amount of water captured in the graphene nanowrinkles is far below the detection limit of techniques capable investigation of the ice formation, such as X-ray or neutron scattering,[17,18] therefore it is almost impossible to gain information about its behavior. The only exception is the Total Internal Reflection Raman spectroscopy, which enables detection of extremely thin layers of water/ice.[19] Nevertheless, the spatial resolution of the TIR Raman spectroscopy is typically in the order of tens of micrometers, which is not sufficient to address the nanoscale feature in the graphene-based samples. Thanks to the presence of the graphene monolayer, we were able to address thermal changes in the confined water *via in situ* Raman spectroscopy *via* the graphene spectral response. The graphene is in fact an atomically thin sensor, which is extremely prone to change its Raman fingerprints induced by variation of doping and strain caused by its interaction with environment. Therefore monitoring the Raman spectra of graphene enables tracking of the temperature-induced phase changes of the water confined between the wrinkled graphene monolayer and the substrate. This strategy is supported by previous studies, in which Raman spectroscopy was used as an optical probe of real-time water diffusion under a graphene monolayer[20] or phase transitions of water confined inside single wall carbon nanotubes.[5]

In our study we collected Raman maps containing ~800 Raman spectra measured between 10 and 370 K over a large sample area (typically 20 x 20 µm$^2$). We succeeded in deconvoluting the signals from the flat and wrinkled graphene regions by careful spectral decomposition of the Raman spectra along with G-2D correlation analysis, according to a procedure previously reported by Lee *et al.*[21] As the lateral dimensions of the wrinkles are below the spatial resolution of the Raman microscope and the nanowrinkles are randomly distributed across the graphene monolayer, the experimental Raman spectra represent a superposition of the contributions from the flat and the wrinkled graphene areas.

The Raman spectra shown in Fig. 1(c) and Fig. 1(d) feature the principal Raman-active modes of graphene: G, 2D and D'.[22] The G and the 2D modes frequencies shift and change their intensity upon strain and doping, which makes them very useful indicators of the band structure variations in the graphene.[22] The frequency shift of the G band in charged graphene is related to the change in the C–C bond strength and to the renormalization of the phonon energy.[23] In the doped (charged) graphene, the Fermi energy $E_F$ is moved away from the Dirac point and thus the formation of electron–hole pairs is suppressed.[23] However, the doping also induces a change of the C–C bond strength.[24] The positive doping removes the electrons from antibonding orbitals and therefore a hardening of the phonon corresponding to the G band is expected. On the other hand, negative doping adds electrons to the antibonding orbitals, which should lead to a softening of the Raman signal frequency ($\omega_G$). Both phonon energy renormalization and a change of the bond strength occur and the two effects are superimposed in the experimental Raman spectra. For positive doping (expected in our sample), both effects lead to an upshift of the phonon frequency. The behavior of the 2D mode is also sensitive to the doping, although the dependence is significantly weaker than in the case of the G mode.[25]

The almost equal integrated intensity of the G and 2D peaks suggest a relatively high doping level of the graphene monolayer.[26] Closer inspection of the figures reveals the asymmetric shape of the G and 2D peaks, which is a typical fingerprint of fine wrinkles and graphene delamination from a substrate. [27]

Thus we decomposed the Raman G and 2D bands into $G_1$, $2D_1$ and $G_2$, $2D_2$ components corresponding to the flat and wrinkled parts of the graphene monolayer in our sample, respectively. It is expected that the flat areas of the graphene do not show significant strain, however they exhibit larger doping due to the proximity of the $SiO_2$/Si substrate, which is well-known source of *p*-type doping.[28] On the other hand the graphene in nanowrinkles is curved therefore the wrinkled graphene reveals substantial strain, but much less doping as the distance from the substrate increases in comparison to the flat graphene.

The $G_1$, $G_2$, $2D_1$, and $2D_2$ components are essential to evaluate the strain and doping levels in the two different types of graphene landscape (flat and wrinkled) in our sample. For that purpose, the intensity (I), Raman shift (ω), and full width at half maximum (FWHM) of all peaks identified in the Raman spectra were determined by fitting pseudo-Voigt functions. Please note that the above mentioned deconvolution is relevant as the roughness of the underlying $SiO_2$/Si substrate is much lower than the dimensions of the topographic features (see the SI, Fig. S1).

Representative maps of the Raman shifts of the $G_1$ (a), $G_2$ (b), $2D_1$ (c), and $2D_2$ (d) peaks obtained at 200 K are shown in Fig. S2. The fitted $G_1$, $G_2$, $2D_1$, and $2D_2$ components were used to evaluate the strain and doping levels in the graphene monolayer.

Using the well-known relations between strain/doping and the Raman shifts described below, we determined the two properties at each point in the G-2D space (Fig. 1(e)), with (1582 cm$^{-1}$, 2674 cm$^{-1}$) as the coordinate origin corresponding to pristine (undoped and unstrained) graphene.

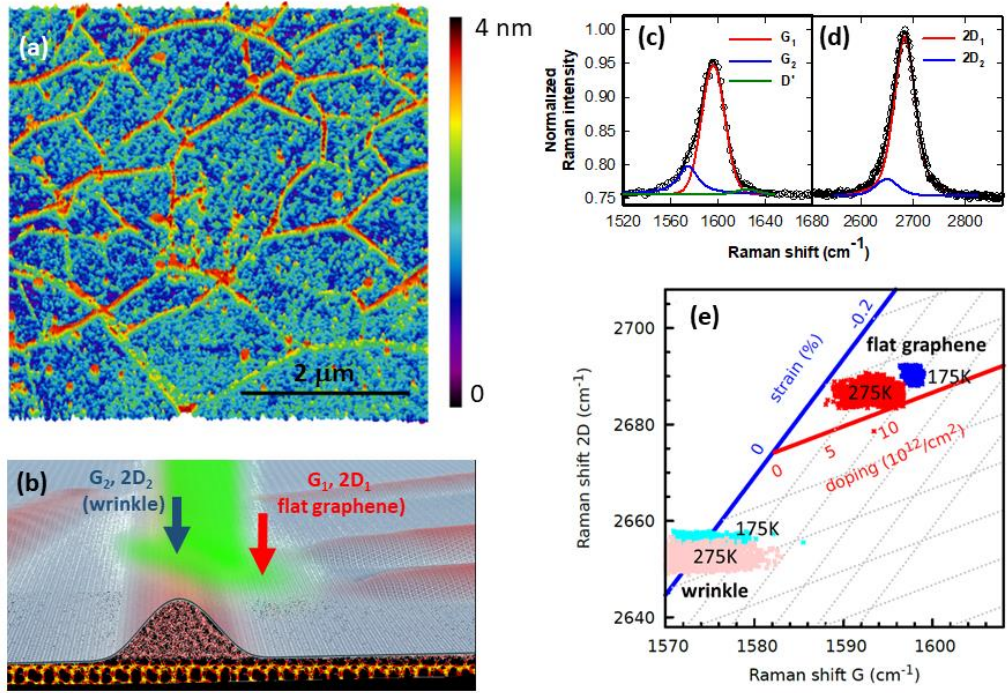

**Figure 1** (a) AFM topography image of graphene monolayer transferred onto the silica substrate *via* the NC-assisted procedure. As described in the text, the monolayer is structured into a network of nanosized wrinkles. (b) Schematic side view of a graphene wrinkle, with water confined between graphene and the silica substrate. The blue and red arrows mark the wrinkled and flat regions, respectively, which can be further resolved by deconvolution of the G and 2D bands of the graphene Raman spectrum, as shown in panels (c) and (d), respectively. In particular, the G (c) and 2D (d) bands can be deconvoluted into contributions originating from the wrinkled (blue lines, $G_2$, $2D_2$) and flat (red lines, $G_1$, $2D_1$) regions of the graphene layer. The minor contribution of the D' peak is shown by the green line in panel (c). The overall Raman spectrum is represented by a solid black line and the experimental data are plotted using open circles. (e) G-2D correlation diagram for the flat and delaminated (wrinkled) areas of the graphene monolayer. The Raman shifts of the $2D_1$ vs $G_1$ ('flat') and $2D_2$ vs $G_2$ ('wrinkle') peaks are grouped into low doping–moderate strain (pink and cyan spots) and medium doping–moderate strain (red and blue spots) areas in the diagram, respectively. The solid blue and red lines correspond to the pure biaxial and pure doping variations, with slopes of 2.45 and 0.7, respectively. Data obtained for temperatures both below (175 K) and above (275 K) the $T_\mathrm{m}$ are included in the diagram.

We calculated the actual strain assuming that graphene was under biaxial strain, with a slope $\Delta\omega_{2D}/\Delta\omega G$ = 2.45 and sensitivity $\Delta\omega G$ = -57 cm$^{-1}$/1%.[29] For calculating the doping contribution, we assumed $\Delta\omega 2D/\Delta\omega G$ = 0.7. The doping-induced shift of $\Delta\omega G$ was calculated according to Das et al.,[25] who reported that the G peak stiffens and sharpens for both electron and hole doping, while the 2D peak shows a different response to holes and electrons. The ratio of the intensities of the G and 2D peaks thus shows a strong dependence on the charge carrier concentration, making it a sensitive parameter to monitor the doping level. To take into account the intrinsic influence of the temperature on the Raman shift of the G band, we further corrected the latter value for temperature-dependent phonon anharmonicity effects.[30] It should be noted that thermal variations always induce a substrate-graphene mismatch, which sometimes may even cause the damage of the graphene. Nevertheless, the temperature invariance of the D band (Fig. S3) points to the absence of any significant harm to our graphene sample.

The correlation diagram displayed in Fig. 1(e) reveals that the (G, 2D) Raman shift points corresponding to the flat and wrinkled graphene regions are grouped into two isolated clusters; in particular, the figure shows the data obtained at 175 and 275 K as representative examples. In agreement with previous studies,[31,32] we unambiguously attribute the pink (275 K) and light blue (175 K) clusters located in the low-doping region to the graphene wrinkles and the red (275 K) and dark blue (175 K) clusters in the high-doping region to the flat graphene.

The temperature-induced variations of the doping and strain levels in the flat and wrinkled graphene areas separately were finally determined at all measured temperatures. The temperature dependence of the doping in the flat and wrinkled graphene areas is shown in Fig. 2(a). A clear kink appears at approximately 200 K for flat graphene, while the doping curve for the wrinkled region only shows a gradual variation with increasing temperature. In contrast, for wrinkled graphene the strain decreases steeply upon cooling down to a temperature of around 200 K (Fig. 2(b)).

The same feature has been observed for the flat graphene in the negative strain scale. The observed trends significantly contrast the behavior of CVD-obtained graphene transferred onto a $SiO_2$/Si substrate using a standard PMMA protocol[33,34] or suspended graphene monolayer.[34] In case of the PMMA transferred graphene, the monolayer follows the thermal expansion of the silicon substrate and clearly reproduces the well-known phonon anomaly in silicon, as we demonstrated in a previous study.[31] Consequently, the temperature dependencies of the strain and the doping in the flat as well as in the wrinkled areas are identical and follow the thermal expansion dependence of the silicon.

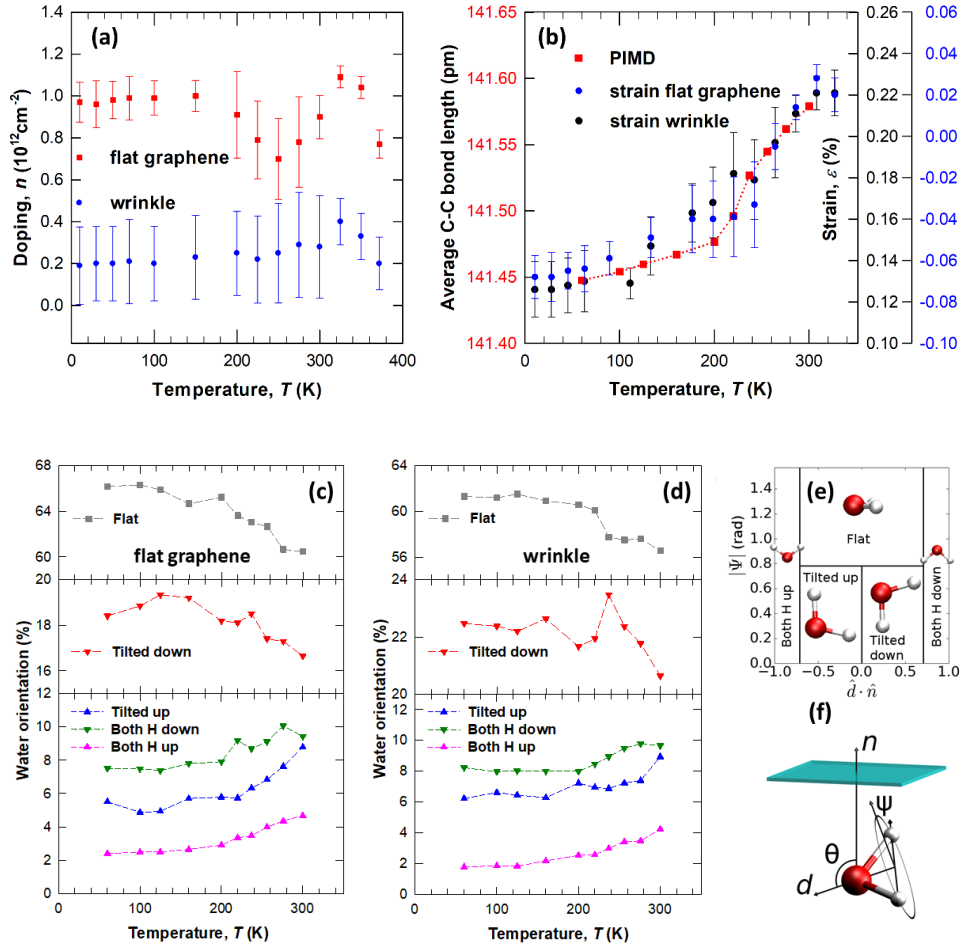

**Figure 2** Temperature dependence of graphene doping and strain levels derived from Raman spectral analysis and of water orientation distributions obtained from the PIMD calculations. (a) Temperature dependence of doping level in flat and wrinkled graphene areas. (b) Comparison of temperature dependence of average C-C bond length and strain levels (derived from the thermal variations of the Raman bands) in wrinkled and flat areas of the graphene monolayer, exhibiting matching trends. (c, d) Fractions of water molecules in different orientations in the flat (c) and wrinkled (d) graphene regions according to the orientations defined in the panel (e) based on the definition of vectors used to obtain the molecular rotational angle $\Psi$ and the projection of the normalized dipole on the graphene normal $\hat{d}\cdot\hat{n}$ given in the panel (f).

The different trends observed for the present system originate from the reorganization of water molecules entrapped below the graphene layer, most likely related to order–disorder transformation and melting processes of the trapped molecules. The observed changes in the properties of graphene, however, occur at temperatures significantly lower than the normal melting temperature of ice. The present experimental results thus point to anomalous changes in the structure of the trapped water molecules during cooling/heating.

To investigate the microscopic mechanism behind the observed effects, we carried out extensive classical and path integral molecular dynamics simulations (MD and PIMD, respectively). The model of the system composed of a silica substrate and a graphene overlayer with wrinkle of height and width of approximately 40Å. Water in the form of hexagonal ice was added between the substrate and the graphene. In the flat region one water layer was present (see visualization of the model in Fig. S5). We note that the exact state of water in the wrinkle at low temperatures is unknown and is difficult to determine due to the low temperatures needed (around 200 K) and large size of the system (over 90 000 atoms).[35,36] The ice Ih model that we use is consistent with water structure found for finite water clusters or confined water.[8,37]

After equilibration at 210 K, the ice structure at the interface with graphene and silica changed into a disordered interfacial layer. The largest change in the water structure occurred at the top of the wrinkled regions, where the graphene layer was highly curved, and in the intermediate regions between wrinkled and flat graphene. Subsequent simulations showed that with increasing temperature, the size of the disordered regions increased and that of the ice core decreased. This behavior is illustrated by the simulation snapshots shown in panels (a), (b), and (c) of Fig. 3, extracted from the MD trajectories at 160, 220, and 256 K, respectively. Complete melting of the ice core occurred between 237 and 246 K, which is approximately 30 K below the melting temperature of the TIP4P/ICE water model ($T_m$ = 270 ± 3 K).[38] This lowering of melting temperature (sometimes termed "supercooling")[11,39,40] is caused by the finite size of water in wrinkle. The observed behavior is in agreement with the work of Limmer and Chandler, who reported a phase diagram of water confined in hydrophilic nanopores of different size and shape.[11]

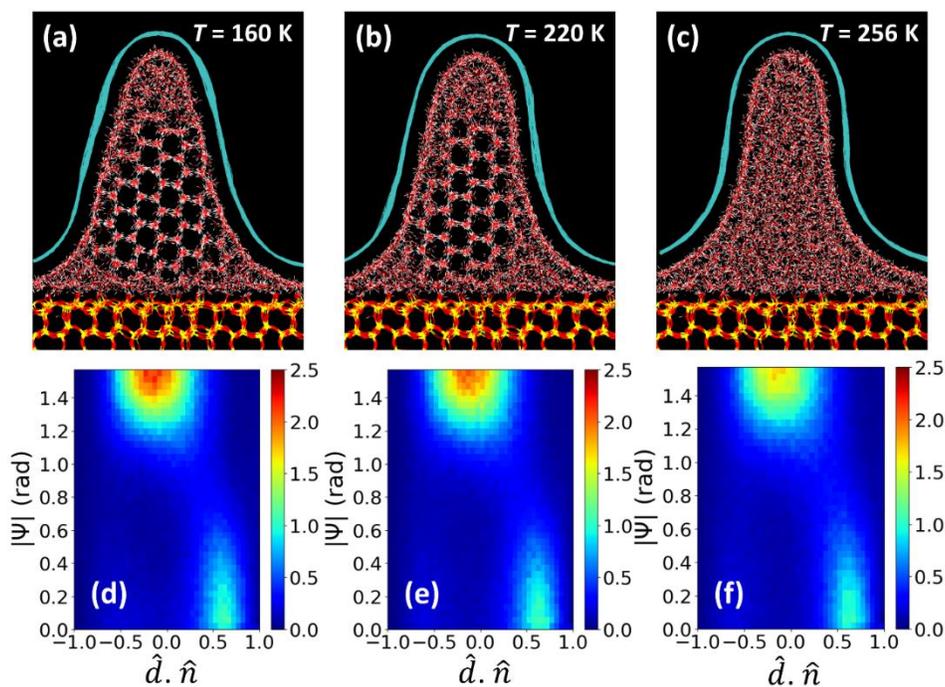

**Figure 3** (a)–(c) Snapshots of water structure within graphene wrinkles at 160 K (a), 220 K (b), and 256 K (c). The oxygen, hydrogen and silicon atoms are given in red, white and yellow, respectively. The graphene monolayer matching the shape of the wrinkle is shown in cyan. (d)–(f) Normalized 2D histograms of water orientation parameters at 160 K (d), 220 K (e), and 256 K (f). The color code indicates the probability distribution of the water orientation according to the two angles. Definition of vectors used to obtain the molecular rotational angle $\Psi$ and the projection of the normalized dipole on the graphene normal $\hat{d}\cdot\hat{n}$ is given in Fig. 2 (c, f).

We analyzed the structure and dynamics of water in contact with the graphene layer to understand how the temperature-dependent changes affect the observed properties of graphene layer.

First we discuss the orientation of water using the water dipole projection on the graphene surface normal and water rotation along the dipole, according to Figs. 2(e) and (f). Panels (d), (e), and (f) of Fig. 3 show histograms of water orientations for water in the interfacial layer for PIMD trajectories at 160, 220 and 256 K. In all cases, there is a clear preference for the water molecules to lie parallel to the graphene surface (flat orientation), with a smaller fraction oriented with one hydrogen atom pointing away from the graphene sheet (tilted down orientation). Clearly, with increasing temperature, the distributions become more blurred, corresponding to increased dynamics at the interface. To examine the structure of water in more detail, we calculated the fractions of molecules in the different orientations defined in Fig. 2(e). The temperature dependence of each fraction is shown in Fig. 2(c) for water molecules in flat graphene regions and in Fig. 2(d) for water molecules in the interfacial layer in wrinkles. The data show significant changes in orientation starting at approximately 200 K for water molecules in both the flat and wrinkled graphene regions. However, for flat regions the drop in "flat" orientation is gradual, see Fig. 2(c), while for the wrinkle regions the drop is more abrupt, see Fig. 2(d). This different behavior corresponds to the different environment of individual water molecules in the flat and wrinkle regions. In the first, water molecules are part of a disordered 2D layer where the potential is softer. In the latter, the molecules are part of ice surface and increased dynamics is possible only above the surface premelting temperature. The increased dynamics of water molecules in the flat regions correlates well with the changes in the doping of graphene.

We note that the orientation of water molecules is not the only property affected by the increasing temperature in our simulations. We observe similar trends, such as a faster increase at temperatures above 200 K (see Fig. S6), also for water self-diffusion coefficient. Overall, the temperature dependences of the calculated and experimental properties corroborate the scenario in which the significant change in doping levels observed *via* graphene probe in the melting region (200 – 270 K) is driven by enhanced water dynamics. This variation in the orientation of water molecules has a counterpart at the macroscopic level, *i.e.*, the change in the dielectric constant of water, along with its charge screening properties, during melting/freezing.

We also compared the temperature dependence of the experimentally measured strain in the graphene wrinkles to that of the C-C bond length in the same regions, obtained from PIMD calculations. The latter approach was needed because standard MD simulations without a quantum description of the nuclei are unable to describe the experimental trend: see Fig. S7. The comparison in Fig. 2(b) shows a perfect match between the experimental and computational data, with a clear kink when the temperature reaches the melting region. Our results indicate that the ice melting and related ice/water volume changes have direct impact on the capping graphene monolayer. Clearly reorganization of water molecules during the melting induces change of the graphene shape, mostly at the wrinkles.

To address the effect of melting on the shape of the graphene more in detail, we have performed PIMD simulations with the water molecules fixed (using the initial structure from the simulation at 160 K). Due to the surface pre-melting, the increased dynamics should not be sensitive to the water phase inside the wrinkle. In this case, the C-C bond length increases with temperature, but there is no anomaly in the temperature dependence as with the free water (Fig. 2 (b)). Also, PIMD simulations of an isolated graphene double-layer did not reveal any change of slope in the temperature dependence of the C-C bond length (see Fig. S8).

The present study thus reveal the interesting interplay between the structures of the wrinkle regions and the structure of water trapped within them. The wrinkle shape defines the volume filled by water, while melting and flowing water in turn affects the shape of the wrinkle. Nevertheless, the results of cryogenic Raman spectroscopy and PIMD simulations points to melting of the water confined in nanoscale graphene wrinkles occurring well below the melting temperature of bulk water.

CONCLUSIONS

In summary, we achieved confinement of water molecules below a nanostructured graphene monolayer deposited on a silica surface. Graphene was used as a highly sensitive probe of the behavior of confined water molecules, using cryogenic Raman spectroscopy to monitor its doping and strain levels. In this way, we observed a clear signature for water molecule reorientation in the Raman spectra of the confining graphene monolayer. The most prominent feature observed in the experiments was a change in the doping level with temperature. Based on the simulations, we explained this change by water melting, which leads to enhanced dynamics and reorientation motions of water molecules. In addition, the temperature dependence of the C-C bond length obtained from PIMD simulations was found to be in excellent agreement with the experimentally derived strain dependences of graphene. Overall, these results allow us to conclude that the water confined within the nanoscale graphene wrinkles undergoes significant reduction of melting temperature, which can be tracked by monitoring the graphene response over realistic time scales being a great benefit in comparison to other techniques routinely used for investigation of phase transitions in confined water.[17,41] Therefore our study introduces an approach for experimental research on confined liquids, as well as on various processes (including chemical reactions) in the regime where liquids are essentially super-cooled. Achieving further control over the properties of confined water is a major challenge, addressing which would provide a gate for entering the 'no man's land' of the water phase diagram at the nanoscale.

METHODS/EXPERIMENTAL

Entrapment of water under graphene was achieved *via* a nitrocellulose-assisted copper etching method.[15,16] First, graphene was grown by CVD on a copper foil, following a previously reported procedure.[42] In brief, the polycrystalline 0.025 mm thick copper foil (Daubert Cromwell) was annealed in temperature of 1000 ºC in the hydrogen (Messer, 6.0 purity) flow of 50 sccm for 20 minutes when the methane precursor was switched on. The graphene growth was carried for 30 minutes with the flow of methane (Messer, 5.0 purity) of 1 sccm. Then the sample was annealed another 5 minutes in $H_2$ to etch the potential add-layers and cooled down to room temperature. The pressure during the whole process was kept at about 0.35 Torr.

The copper foil with as-grown graphene was covered with a 50 nm-thick nitrocellulose film by spin-coating a 2% solution of nitrocellulose in amyl acetate (Sigma Aldrich, #09817). The copper foil was etched away using copper etchant type CE-100 (Transene) composed of Ferric

Chloride (25-35 wt %), Hydrochloric Acid (3-4 wt %) and water (> 60 wt %) with etching time 5 – 10 minutes, after which the sample was 'fished' using a Si/SiO$_2$ wafer (300 nm thermal SiO$_2$ film on Si, SQI) and subsequently washed several times using water purified by NANOpure system. Nitrocellulose film was removed by flushing the sample several times with methanol (Methanol Micropore Ulsi, Technic). The samples were finally dried with nitrogen gas (Messer, 6.0).

To demonstrate different nature of the graphene obtained by the nitrocellulose-assisted transfer monolayer, which enables entrapping the water molecules, the sample was subjected to an annealing test (150 °C for 1 h) and then investigated by Raman spectroscopy (Fig. S4).

AFM measurements were carried out using PeakForce quantitative nanomechanical mapping (QNM) and imaging in tapping mode, using a Dimension Icon microscope (Bruker Inc.). Images were captured using Bruker Scanasyst-Air probes (k = 0.4 N/m, f0 =70 kHz, nominal tip radius = 2 nm). High-quality images were processed in the standard way using the Gwyddion software,[43] applying line-by-line 1st-order leveling and scar removal.

Low-temperature Raman spectroscopy experiments were carried out using a confocal microscope (attoRAMAN, Attocube) placed in a physical property measurement system (Quantum Design). After mounting the sample onto the microscope, the sample space was flushed several times with He gas and then evacuated to 5 mbar. RS were acquired using a WITec Alpha300 spectrometer with 2.32 eV (532 nm) laser excitation, a grating of 600 lines mm$^{-1}$, and a 100× objective (numerical aperture = 0.82). The sample temperature was monitored with a thermometer placed directly under it. 2D Raman maps of $20 \times 20$ μm$^2$ size were acquired with lateral steps of 500 nm in both directions, focusing a maximum laser power of 1 mW on the sample. The resolution of the Raman microscope is ~ 1 μm (500 nm nominal) and the acquisition time was 5 s.

MD simulations were carried out using the GROMACS 5.1.4.[44] and LAMMPS[45] software packages. The silica substrate was modeled using the structure and force field developed by Emami *et al*.[46] In MD simulations we used the TIP4P/ICE model for water, for carbon atoms we used the AMBER force field parametrization of aromatic carbon. In PIMD[47] simulations we used the q-TIP4P/F model for water and the AIREBO potential for graphene.[48] Interaction parameters not present in the above force fields were set according to the Lorentz–Berthelot mixing rules.

The PIMD runs were performed using LAMMPS with an i-PI driver.[49] The path integral generalized Langevin equation thermostat (PIGLET)[50] was used to reduce the number of replicas required for the convergence of the PIMD simulations. The initial configuration of the model was generated by matching the experimental geometry of the graphene wrinkles. The simulation cell sizes along the $x$ and $y$ directions were $133.468 \times 209.052$ Å, and the cell contained more than 90,000 atoms (see visualization of the model in the Fig. S5). The initial height and width of the simulated wrinkle were approximately 40 Å, its initial shape was a scaled cosine function from an interval between $-\pi$ and $\pi$. A flat graphene was added to the wrinkle to create a continuous layer.

The initial structure of water was that of hexagonal ice. After initial equilibration at low temperatures, production runs for temperatures 210 K and higher were performed for 5.25 ns. For lower temperatures the 210 K structure was used and run for 2.25 ns. For each temperature the structures from the last 0.75 ns were used to make data in Fig. 2 and Fig. 3. The final structures were used as input to LAMMPS to run equilibration with AIREBO and TIP4P/ICE models for at least 225 ps. The resulting structures were used as input for PIMD calculations with q-TIP4P/F and AIREBO models in LAMMPS. Further simulation details can be found in the SI (including Figs. S5-S8 and Tabs. SI and SII).

For analysis, we calculated the orientation of water molecules with respect to the nearest part of the graphene surface. We describe the orientation using the tilt of water from surface normal and rotation of hydrogens. Specifically the tilt is obtained as a dot product of normalized water dipole vector $\hat{d}$ and local normal vector of the graphene surface $\hat{n}$. The rotation $\Psi$ gives the rotation of water molecule around its dipole vector from structure, where one of the hydrogens is closest to the graphene surface, see Fig. 2(e) and (f). Thus $\Psi$ can assume values between 0 and $\pi$, for which the hydrogens are in the same distance to graphene. We used $\hat{d}\cdot\hat{n}$ and $\Psi$ to divide the molecules into five different groups based on their orientation: tilted up, tilted down, both hydrogen atoms up, both hydrogen atoms down, and lying flat. The relations among the groups and the values of $\hat{d}\cdot\hat{n}$ and $\Psi$ are illustrated in Fig. 2(f), and described in detail in Tab. SII.

## ASSOCIATED CONTENT

### SUPPORTING INFORMATION

The Supporting Information is available free of charge at (*actual link will be given here*). Additional results of Raman micro-spectroscopy, AFM images of the substrate, Construction of initial structures, Analysis of molecular dynamics (PDF).

## AUTHOR INFORMATION

### CORRESPONDING AUTHORS

Martin Kalbac, email: martin.kalbac@jh-inst.cas.cz

Jana Vejpravova, email: jana@mag.mff.cuni.cz21For analysis, we calculated the orientation of water molecules with respect to the nearest part of the graphene surface. We describe the orientation using the tilt of water from surface normal and rotation of hydrogens. Specifically the tilt is obtained as a dot product of normalized water dipole vector $\hat{d}$ and local normal vector of the graphene surface $\hat{n}$. The rotation $\Psi$ gives the rotation of water molecule around its dipole vector from structure, where one of the hydrogens is closest to the graphene surface, see Fig. 2(e) and (f). Thus $\Psi$ can assume values between 0 and $\pi$, for which the hydrogens are in the same distance to graphene. We used $\hat{d}\cdot\hat{n}$ and $\Psi$ to divide the molecules into five different groups based on their orientation: tilted up, tilted down, both hydrogen atoms up, both hydrogen atoms down, and lying flat. The relations among the groups and the values of $\hat{d}\cdot\hat{n}$ and $\Psi$ are illustrated in Fig. 2(f), and described in detail in Tab. SII.

## ASSOCIATED CONTENT

### SUPPORTING INFORMATION

The Supporting Information is available free of charge at (*actual link will be given here*). Additional results of Raman micro-spectroscopy, AFM images of the substrate, Construction of initial structures, Analysis of molecular dynamics (PDF).

## AUTHOR INFORMATION

### CORRESPONDING AUTHORS

Martin Kalbac, email: martin.kalbac@jh-inst.cas.cz
Martin Kalbac, email: martin.kalbac@jh-inst.cas.cz

Jana Vejpravova, email: jana@mag.mff.cuni.cz

The authors declare no competing financial interests.

ACKNOWLEDGEMENT

We acknowledge support from the European Research Council, grants TSuNAMI (no. 716265) and APES (no. 759721). JK was also supported by a PRIMUS project from Charles University, Czech Republic. JV, TV and MK would like to acknowledge CSF (Grant no. 18-20357S). Computational resources were provided by the IT4Innovations National Supercomputing Center (LM2015070), CESNET (LM2015042), CERIT-SC (LM2015085), and LM2018140. Low temperature measurements were carried out in Materials Growth and Measurement Laboratory (LM2018096), which are all supported by Ministry of Education, Youth and Sports, Czech Republic.